\documentclass[12pt,preprint,usenatbib]{aastex}

\def\araa{{\rm ARA$\,$\&$\,$A}}                % Annu. Rev. A \and\ A
\def\apj{{\rm ApJ}}                    % Astrophys. J.
\def\apjs{{\rm ApJS}}                  % Astrophys. J. Supl.\def\apjl{{\rm ApJ Let}}               % Astrophys. J. (Letters)
\def\aj{{\rm AJ}}                      % Astron. J.
\def\mnras{{\rm MNRAS}}                        % Monthly Notices.

\def\a4{\hsize 17.0cm \vsize 25.cm}

\shorttitle{Broadened Main Sequences in young massive clusters}
\shortauthors{ Li et al.}

\begin{document}

\title{Discovery of extended main sequence turn-offs in four young massive clusters in the Magellanic Clouds}
\author{Chengyuan Li}
\affil{Department of Physics and Astronomy, Macquarie University, Sydney, NSW 2109, Australia}
\author{Richard de Grijs}
\affil{Kavli Institute for Astronomy \& Astrophysics and Department of Astronomy, Peking University, Yi He Yuan Lu 5, Hai Dian District, Beijing 100871, China}
\affil{International Space Science Institute--Beijing, 1 Nanertiao, Zhongguancun, Hai Dian District, Beijing 100190, China}
\author{Licai Deng}
\affil{Key Laboratory for Optical Astronomy, National Astronomical Observatories, Chinese Academy of Sciences, 20A Datun Road, Chaoyang District, Beijing 100012, China}
\author{Antonino P. Milone}
\affil{Research School of Astronomy \& Astrophysics, Australian National University, Mt Stromlo Observatory, Cotter Road, Weston, ACT 2611, Australia}

\begin{abstract}
An increasing number of young massive clusters (YMCs) in the
Magellanic Clouds have been found to exhibit bimodal or extended main
sequences (MSs) in their color--magnitude diagrams (CMDs). These
features are usually interpreted in terms of a coeval stellar
population with different stellar rotational rates, where the blue and
red MS stars are populated by non- (or slowly) and rapidly rotating
stellar populations, respectively. However, some studies have shown
that an age spread of several million years is required to reproduce
the observed wide turn-off regions in some YMCs. Here we present the
ultraviolet--visual CMDs of four Large and Small Magellanic Cloud
YMCs, NGC 330, NGC 1805, NGC 1818, and NGC 2164, based on
high-precision Hubble Space Telescope photometry. We show that
they all exhibit extended main-sequence turn-offs (MSTOs). The 
importance of age spreads and stellar rotation in reproducing 
the observations is investigated. The observed extended MSTOs cannot 
be explained by stellar rotation alone. Adopting an age spread of 
35--50 Myr can alleviate this difficulty. We conclude that stars in 
these clusters are characterized by ranges in both their ages 
and rotation properties, but the origin of the age spread in 
these clusters remains unknown.
\end{abstract}

  
\section{Introduction}\label{S1}
The origin of the multiple stellar populations that have been
discovered in almost all Galactic globular clusters (GCs) remains an
open question \citep{Grat04a,Piot15a}. Since it is not possible to
observe the young GCs in the early Universe, studying young GC
candidates in the Milky Way and its satellites therefore plays a
fundamental role in addressing this issue. However, almost all young
Galactic clusters are located in the Galactic disk. They are affected
by severe foreground extinction, which may mask the multiple stellar
populations in their observed color--magnitude diagrams (CMDs).

An alternative way is to study the young clusters in the Large and
Small Magellanic Clouds (LMC and SMC). Current observations of star
clusters in the LMC and SMC mainly reveal two features: (a) almost all
intermediate-age (1--2 Gyr-old) star clusters exhibit extended main
sequence turn-offs \citep[eMSTOs; e.g.,][]{Bert03a,Mack07a,
Milo09a,Gira13a,Li14a}, while some even exhibit extended or dual 
red clumps \citep{Gira09a,Li16a}; (b) some young massive clusters 
(YMCs; with ages $\leq$400 Myr) in the LMC exhibit bimodal or 
extended main sequences (MSs) \citep{Milo15a,Bast16a,Milo16a,Milo16b}.

The discovery of eMSTO regions in intermediate-age clusters has
triggered a surge of interest in explaining this feature, starting
from the age-spread scenario, which attributes the observed eMSTOs to
primodial age spreads of up to 700 Myr \citep{Goud11a,Gira13a,Goud14a,
Gira16a}. Next came the rapid stellar rotation scenario
\citep{Bast09a,Yang13a,Li14b,Bran15a,Bast15a,Li16a,Wu16a}. Recently, a
new scenario has been suggested, i.e., that stellar variability may
play a potential role in shaping the eMSTO regions as well
\citep{Sali16a,grijs17a}. The debate is indeed heating up.

Further excitement was added by the discovery of bimodal or extended
MSTOs in YMCs. These features are usually explained by a coeval stellar
population with a dispersion in stellar rotation rates
\citep{Li14a,Li14b,Bast16a} or a coeval, rapidly rotating
population combined with a slowly/non-rotating population exhibiting
an age spread \citep{Milo16b}. It has been suggested that the blue MS
stars may hide a binary component, where binary synchronization is
responsible for their small rotational rates \citep{Dant15a}. A recent
study based on the YMCs NGC 1866 and NGC 1850 showed that a coeval
stellar population featuring a distribution of stellar rotation rates
can only partially explain the broadening of the MSTO \citep{Milo16b,Corr16a}.

In this paper, we study four YMCs in the LMC and SMC, i.e., NGC 330
(SMC), NGC 1805, NGC 1818, and NGC 2164. Their isochronal ages do not
exceed 40 Myr, 40 Myr, 35 Myr, and 100 Myr, respectively. Most are
younger than the recently studied YMCs
\citep[e.g.,][]{Li16b,Milo15a,Milo16a,Milo16b}. We analyzed their CMDs
using high-resolution {\it Hubble Space Telescope} ({\sl HST})
photometry, obtained using ultraviolet and visual filters. We find
that they all exhibit extended MSTOs, although at different levels of
significance. We test whether the observations can be reproduced
assuming an age spread, a dispersion in stellar rotation rates, or a
combination of these scenarios.

This article is arranged as follows. Section \ref{S2} presents our
data reduction, which is followed by the main results of our analysis
in Section \ref{S3}. In Section \ref{S4} we present a
discussion. Section \ref{S5} summarizes and concludes the paper.

\section{Data Reduction}\label{S2}

The data used in this work come from the {\sl HST} Ultraviolet and
Visual Channel of the Wide Field Channel 3 (UVIS/WFC3) and Wide Field
and Planetary Camera 2 (WFPC2) images. The WFC3/UVIS images were
observed through the F225W and F336W passbands and the {\sl HST}/WFP2
images were observed in the F555W and F814W bands. Relevant
information pertaining to the data is summarized in Table \ref{T1}.

\begin{deluxetable*}{llllll}
\tablecaption{Inventory of the data set used in this work\label{T1}}
\tablehead{
Cluster & Camera & exposure time & filter &
program ID & PI name
}
\startdata
NGC 330 & WFC3/UVIS & 10 s + 100 s + 805 s + 3$\times$960 s & F336W & GO-13727 &  J. Kalirai \\
  & WFPC2 & 10 s + 4$\times$350 s & F555W & GO-8134 & A. Nota  \\
NGC 1805 & WFC3/UVIS & 10 s + 100 s + 790 s + 3$\times$947 s & F336W & GO-13727 &  J. Kalirai \\
  & WFPC2 & 3$\times$5 s + 3$\times$140 s + 2$\times$800 s + 900 s & F555W & GO-7307 & G. Gilmore  \\
NGC 1818  & WFC3/UVIS & 10 s + 100 s + 790 s + 3$\times$947 s & F336W & GO-13727 &  J. Kalirai \\
  & WFPC2 & 3$\times$5 s + 3$\times$140 s + 2$\times$800 s + 900 s & F555W & GO-7307 & G. Gilmore  \\
NGC 2164  & WFC3/UVIS & 10 s + 100 s + 790 s + 3$\times$947 s & F336W & GO-13727 &  J. Kalirai \\
  & WFPC2 & 10 s + 4$\times$350 s & F555W & GO-8134 & A. Nota  
\enddata
\end{deluxetable*}

As regards the WFC3/UVIS and WFPC2 images, we performed
point-spread-function (PSF) photometry on the `\_flt' or `\_c0m'
frames using the WFC3 and WFPC2 module of the {\sc dolphot 2.0}
photometry package \citep{Dolp11a,Dolp11b,Dolp13a}. To obtain a
stellar catalog with high-quality photometry, we adopted a filter
employing the sharpness and `crowding' parameters calculated by {\sc
  dolphot}. The sharpness is a measure comparing the profile of an
object relative to the PSF. A perfect star would have a sharpness of
zero. A negative sharpness that is too small usually indicates a
cosmic ray, while a very large positive sharpness means that the
detected object is extended, probably a background galaxy. The
crowding parameter quantifies how much brighter a star would have been
measured had nearby stars not been fitted simultaneously (in units of
magnitudes). For an isolated star, the crowding is zero. High crowding
usually means that the star is poorly measured. We confirmed that most
high-crowding objects are faint, which would not dramatically affect
our analysis. We only selected objects with
$-$0.2$\leq$sharpness$\leq$0.2 and crowding$\leq$0.5 in both frames,
which left us with $\sim$70\% of the objects with high-accuracy
photometry. The number of objects before and after the selection (in
brackets) for the clusters NGC 330, NGC 1805, NGC 1818, and NGC 2164
were 9996 (7208), 3808 (3102), 4478 (3346), and 4284 (3322),
respectively. {\sc dolphot} can automatically flag `good stars' and
centrally saturated objects. We only kept objects that were indicated
as good in our final sample for further analysis.

For each cluster, we combined stellar catalogs of different exposure
times into a deep catalog. If one star appeared in two stellar
catalogs, we selected the longest exposure time as best
representation. Next, we obtained two stellar catalogs, derived from
the WFC3/UVIS and WFPC2 images. We transferred each star's CCD
coordinates to equatorial coordinates ($X,Y$ $\rightarrow$
$\alpha_{\rm J2000}$, $\delta_{\rm J2000}$). We only selected stars
located in the areas covered by both CCD fields for each cluster. This
process sacrifices a large field but provides us with deep, multi-band
stellar catalogs.

We used the method introduced by \citet[][their Section 3.1]{Milo12a}
to calculate differential reddening maps. We corrected all stars for
differential reddening by assuming $A_{\rm UVIS,F336W}=1.658A_{\rm V}$
and $A_{\rm WFPC2,F555W}=1.017A_{\rm V}$. We found that differential
reddening does not dramatically change the morphology of our CMDs.

\begin{figure*}
\centering
\includegraphics[width=0.85\textwidth]{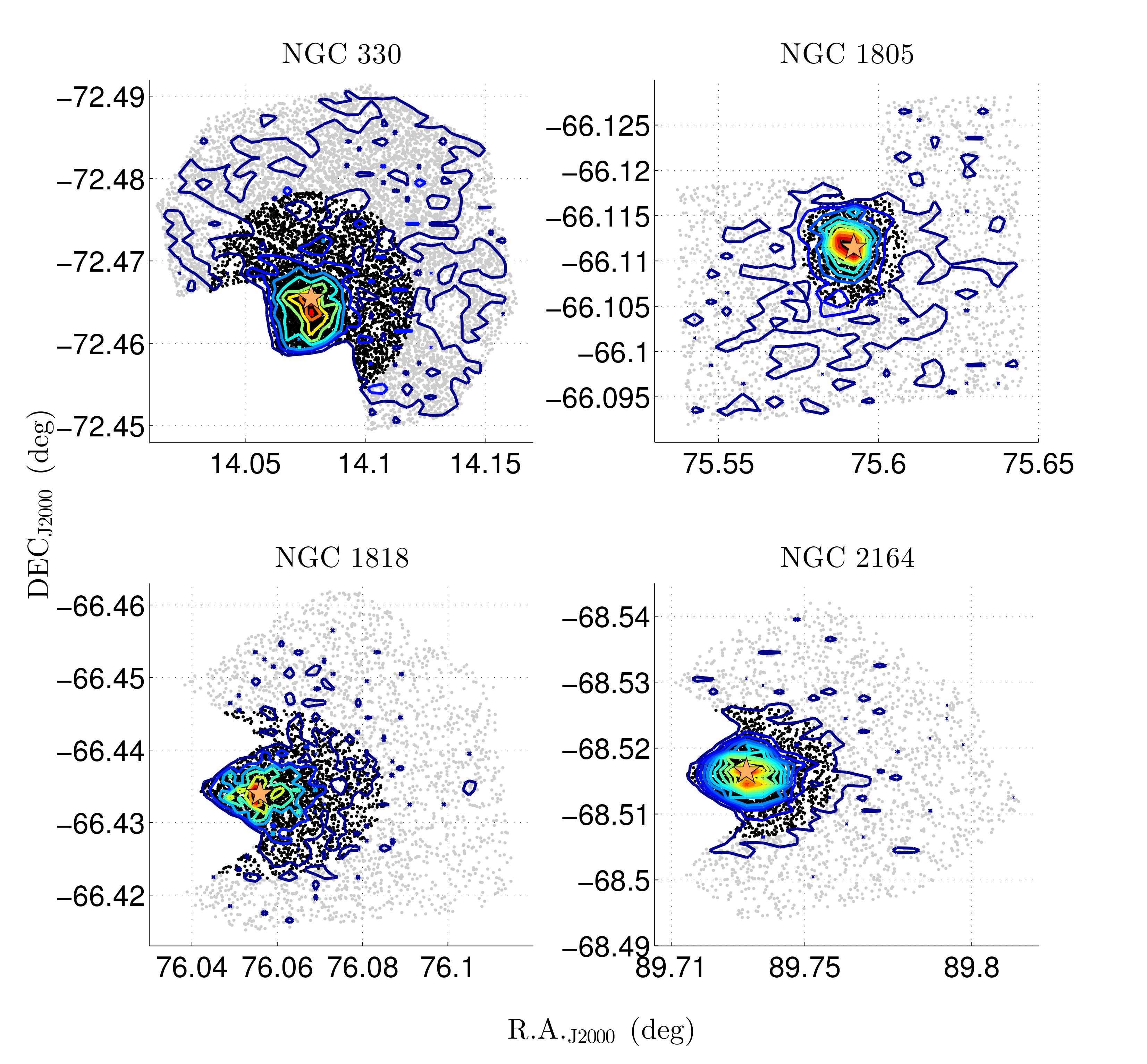}
\caption{Spatial distributions of all stars (gray dots) and sample
  stars within two half-light radii (black dots). The cluster centers
  are indicated by orange pentagrams. The corresponding number-density
  contours are included. Since the field fluctuations are very small
  compared to the central density levels, for NGC 1818 and NGC 2164
  there are no apparent density contours in the outer fields.}
\label{F1}
\end{figure*}

We generated number-density contour figures to determine the center
coordinates of our clusters. We simply assigned the coordinate where
the number density reaches the highest value as the cluster center. We
then used the number-density center to define a circular region within
twice the half-light radius (2$r_{\rm hl}$) as the cluster region. The
$r_{\rm hl}$ values were taken from \cite{Mcla05a}, which were based
on fits to a Wilson model \citep{Wils75a}\footnote{\cite{Wils75a}
  models are spherical and isotropic versions of models usually
  applied to elliptical galaxies. \cite{Mcla05a} found that for
  $\sim90$\% of their full sample of YMCs and old GCs, \cite{Wils75a}
  models provide equally good or significantly better fits than
  \cite{King66a} models}. In Fig. \ref{F1} we show the stellar spatial
distributions, as well as their corresponding cluster centers and
number-density contours. We confirmed that for all clusters, the
stellar number density decreased to the field level at 2$r_{\rm
  hl}$. The clusters' structural parameters are included in Table
\ref{T2}.

\begin{deluxetable*}{llllll}
\tablecaption{Structural parameters \label{T2}}
\tablehead{
Cluster & $\alpha_{\rm J2000}$ & $\delta_{\rm J2000}$ & $h_{\rm c}$ (pc) & $h_{\rm hl}$ (pc)&\\
 & & & (1) & (2) &
}
\startdata
NGC 330 & $00^{\rm h}56^{\rm m}31.00^{\rm s}$  & $-72^{\circ}27'55.80''$ & 2.47$^{+0.14}_{-0.19}$  & 6.95$^{+0.96}_{-0.43}$ \\
NGC 1805 & $05^{\rm h}02^{\rm m}22.14^{\rm s}$  & $-66^{\circ}06'41.40''$ & 1.34$^{+0.09}_{-0.10}$  & 2.86$^{+0.13}_{-0.08}$\\
NGC 1818 & $05^{\rm h}04^{\rm m}13.44^{\rm s}$  & $-66^{\circ}26'02.40''$ & 2.36$^{+0.08}_{-0.09}$  & 5.05$\pm0.08$\\
NGC 2164 & $05^{\rm h}58^{\rm m}55.80^{\rm s}$  & $-68^{\circ}30'59.40''$ & 1.69$^{+0.09}_{-0.05}$  & 4.37$^{+0.05}_{-0.09}$
\enddata
\tablecomments{
(1), (2): \cite{Mcla05a}
}
\end{deluxetable*}

The final step is field-star decontamination. For NGC 330, NGC 1805,
NGC 1818, and NGC 2164, we adopt stars that are located beyond
3.75$r_{\rm hl}$, 6.00$r_{\rm hl}$, 4.00$r_{\rm hl}$, and 4.00$r_{\rm
  hl}$, respectively, as our reference field stars. This choice
results in numbers of field stars close to $\sim$500 for all
clusters. We decontaminated the cluster CMDs using a similar method to
that employed by \cite{Li13a}. We divided both the cluster and field
CMDs into a carefully considered number of cells and counted the
number of stars in each. The adopted numbers of grid cells were used
to determine a reasonable average size for the grid cells. They should
contain enough field stars to statistically correct the cluster CMDs,
but they should not be too large, since that would prevent us from
distinguishing detailed features along the MSs. Finally we randomly
removed stars corresponding to those in the area-corrected field-star
CMDs from the cluster CMDs. We found that our method tends to
oversubtract field stars in the faint tail of the MS. This is because
the field regions usually have lower detection limits than the
clusters' central regions because of crowding and background effects
\cite[which results in higher completeness levels for the field
  samples at the bottom of the MSs; see Fig. 7 of][]{Li13a}. In this
paper, we only focus on the MS range covering F555W$\leq$21 mag, which
is $\sim$3 mag brighter than the magnitude of the detection
limits. Our field-star contamination is thus reliable for the bright
MSs.

\section{Main Results}\label{S3}
\begin{figure*}
\centering
\includegraphics[width=0.85\textwidth]{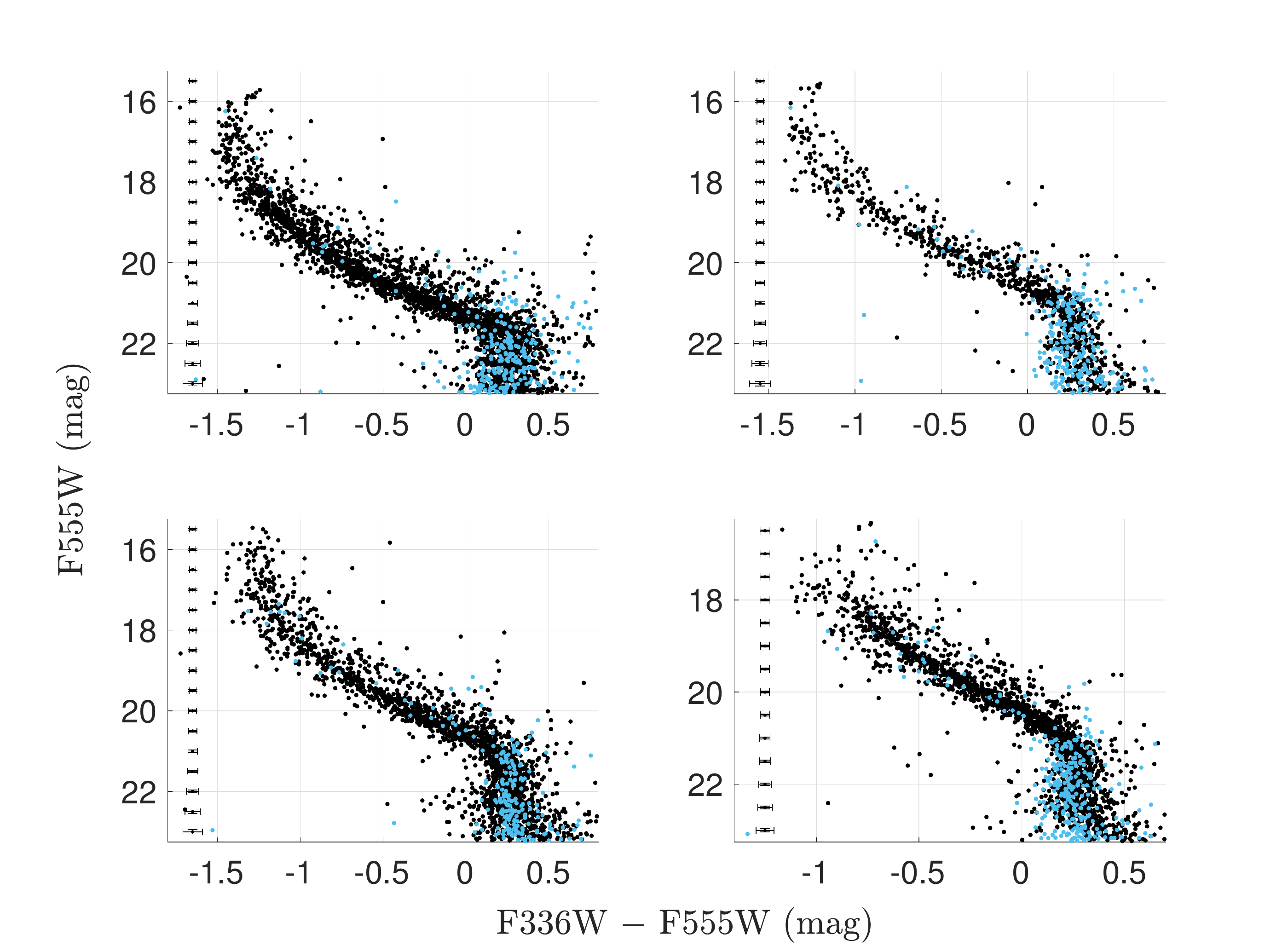}
\caption{Decontaminated CMDs of NGC 330, NGC 1805, NGC 1818, and NGC
  2164. The reference field stars (blue circles) are also shown.
    The corresponding average (1$\sigma$) photometric uncertainties
    are shown on the left-hand sides of all panels.}
\label{F2}
\end{figure*}

The decontaminated YMC CMDs are presented in Fig. \ref{F2}. Here we
only present the (F555W versus F336W--F555W) CMDs, because the
broadening of the MS is most obvious in this parameter
space. Nevertheless, this feature can be detected in other
ultraviolet--visual CMDs as well, e.g., in the (F814W versus
F336W--F814W) CMDs. As shown in Fig. \ref{F2}, all clusters exhibit
wide MSs. We show the distribution of reference field stars in the
same CMD. Because the clusters' broadened MS regions are bright,
contamination by field stars is minimal.

We next used different isochrones to fit the
observations. Specifically, we used two different stellar evolution
models to compre with the observations, i.e., the Geneva SYCLIST code
\citep{Ekst12a,Geor13a,Geor14a}\footnote{\url
  {https://obswww.unige.ch/Recherche/evoldb/index/}} and the MESA
Isochrone \& Stellar Tracks (MIST) models
\citep{Paxt11a,Paxt13a,Paxt15a,Dott16a,Choi16a}\footnote{\url
  {http://waps.cfa.harvard.edu/MIST/}}. We first used the Geneva
SYCLIST database to test if a coeval population of stars characterized
by different rotation rates could reproduce the observations. We
overplotted nine Geneva isochrones with rotation rates of
  $\omega$ = 0.0--0.95\footnote{$\omega$ = 0.0, 0.1, 0.3, 0.5,
    0.6, 0.7, 0.8, 0.9 and 0.95} to the observed CMD, where $\omega$
represent the ratio of the stellar rotation rate to the critical,
breakup rate. All isochrones have the same age. The effects of
extinction are not included in the Geneva models; we simply shifted
the isochrones to match the MS ridge-lines. The input metallicities in
the Geneva models are limited to $Z$ = 0.002, 0.006, and 0.014
($Z_{\odot}$). The physical parameters adopted for our fits are
included in Table \ref{T3} (rows 1--4). Our fits were based on initial
estimates taken from previous publications. The typical age of NGC 330
is 20 Myr or $\log( t \mbox{ yr}^{-1})= 7.30$
\citep{Mcla05a}. For NGC 1805 and NGC 1818, the relevant numbers are
25 Myr to 40 Myr, i.e., $\log( t \mbox{ yr}^{-1})=7.40$--7.60
\citep{John01a}, and for NGC 2164 it is 80 Myr, $\log( t \mbox{
  yr}^{-1})=7.91$ \citep{Mucc06a}. The best-fitting metallicities are
      [Fe/H] = $-$0.82 dex \citep[0.15$Z_{\odot}$;][]{Mcla05a}, $-$0.3
      dex \cite[$0.5Z_{\odot}$;][]{Li13a}, 0.0 dex
      \citep[$Z_{\odot}$;][]{Li13a}, and $-$0.4 -- 0.0 dex
      \citep[0.4$Z_{\odot}$ to $Z_{\odot}$;][]{Saga91a} for NGC 330,
      NGC 1805, NGC 1818, and NGC 2164, respectively. The extinction,
      $E(B-V)$, for these clusters varies from 0.08 mag to 0.12 mag
      \citep{Bess91a,Li13a,Vall91a}. The canonical distance moduli to
      the LMC and SMC are $(m-M)_0$ = 18.49 mag and 18.96 mag,
      respectively \citep{grijs14a,grijs15a}.

\begin{deluxetable*}{llllllll}
\tablecaption{Physical parameters for the isochrones or synthetic clusters in Figs \ref{F3} and \ref{F4} (rows 1--4), Fig. \ref{F5} (rows 5--8), and Figs \ref{F6}--\ref{F9} (rows 9--12). \label{T3}}
\tablehead{
Cluster & $\log( t \mbox{ yr}^{-1})$ & $Z$ & $E(B-V)$ (mag) & $(m-M)_0$ (mag) & $\omega$}
\startdata
NGC 330 & 7.30 & 0.002 &N/A & 19.00 & 0.00--0.95\\
NGC 1805 & 7.50 & 0.006 &N/A & 18.50 & 0.00--0.95 \\
NGC 1818 & 7.55 & 0.014 &N/A & 18.55 & 0.00--0.95 \\
NGC 2164 & 8.00 & 0.014 &N/A & 18.55 & 0.00--0.95\\
\hline
NGC 330 & 6.00,7.40,7.60 & 0.003 & 0.10 & 18.90  & 0.40\\
NGC 1805 & 6.00,7.30,7.60 & 0.006 & 0.12 & 18.40  & 0.40 \\
NGC 1818 & 6.00,7.25,7.55 & 0.011 & 0.08 & 18.60  & 0.40\\
NGC 2164 & 7.70,7.85,8.00 & 0.011 & 0.10 & 18.55 & 0.40\\
\hline
NGC 330 & 6.00--7.60 & 0.002 & N/A & 19.00 & 0.00--0.95\\
NGC 1805 & 6.00--7.60 & 0.006 & N/A & 18.50 & 0.00--0.95 \\
NGC 1818 & 6.00--7.55 & 0.014 & N/A & 18.55 & 0.00--0.95\\
NGC 2164 & 7.70--8.00 & 0.014 & N/A & 18.55 & 0.00--0.95
\enddata
\end{deluxetable*}

\begin{figure*}
\centering
\includegraphics[width=0.85\textwidth]{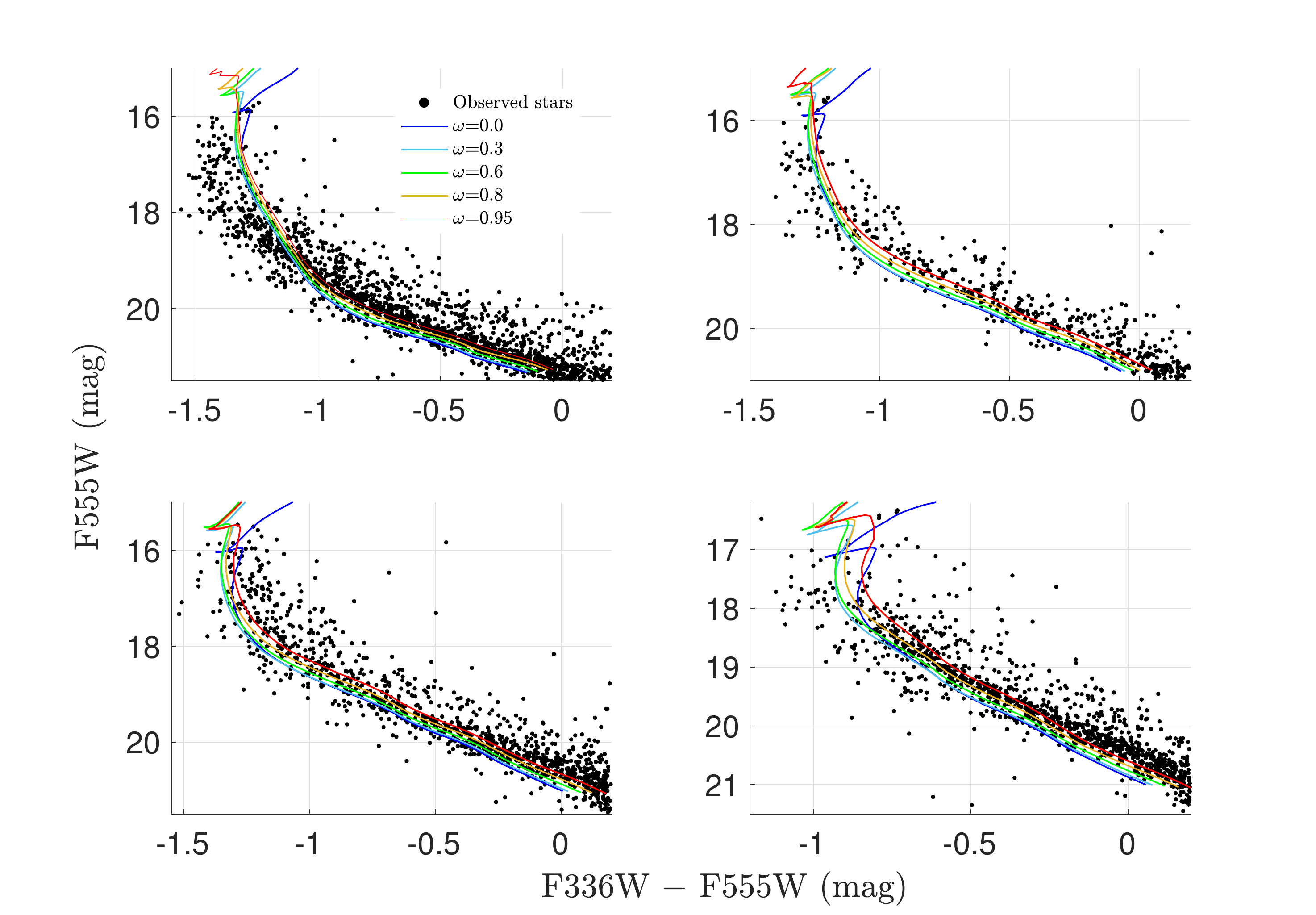}
\caption{Observed CMDs and best-fitting isochrones based on the Geneva
  SYCLIST code. The solid lines of different colors are
    isochrones characterized by different rotation rates, as indicated
    in the legend. For reasons of clarity, we only display five
    isochrones with different rotation rates.}
\label{F3}
\end{figure*}

We show our fits in Fig. \ref{F3}.We find that for three young
  star clusters (NGC 330, NGC 1805 and NGC 1818), the non- and
  fast-rotating isochrones converge in the MSTO region. Although
  isochrones with different rotation rates create a broadened region
  near the MSTO region of NGC 2164, the broadening still seems too
  narrow to fully explain its extended MSTO region. For all our
clusters, the observed MS widths gradually increase toward the MSTO
region. Our fits seem to show minor disagreements between
the models and the observations.

Isochrone fitting alone cannot precisely describe the similarities
between the models and the observations. In order to better
demonstrate the reliability of our fits, for each cluster we
constructed a synthetic CMD for comparison. Our synthetic CMDs are
based on the adopted isochrones, to which we added the same
photometric uncertainties as pertaining to the real data. We used the
method of \cite{Milo12a}. We determined MS--MS binary fractions of
$53.8\pm4.9$\%, $46.1\pm3.5$\%, $39.9\pm2.9$\%, and $53.3\pm4.5$\% for
NGC 330, NGC 1805, NGC 1818, and NGC 2164, respectively\footnote{These
  are the binary fractions for all mass ratios, under the assumption
  that the mass-ratio distribution is flat.}. The average binary
fraction of these four clusters is $\sim$48.3\%. We thus include 50\%
of unresolved MS--MS binaries in each simulated CMD. If the mass
  of the secondary component of a simulated binary system below
  1.7$M_{\odot}$ (corresponding to the low-mass limit of the grid of
  B-type stars with different rotation rates in the Geneva models), we
  interpolate the magnitudes of the secondary star using the
  isochrones calculated based on the large grid
  \citep[$M=0.8$--$500M_{\odot}$, only two rotation
    rates;][]{Ekst12a,Geor13a}. Where necessary, we extrapolate
  outside the grid's boundaries down to 0.08 $M_{\odot}$ (i.e., the
  stellar hydrogen-burning limit).

\begin{figure*}
\centering \includegraphics[width=0.8\textwidth]{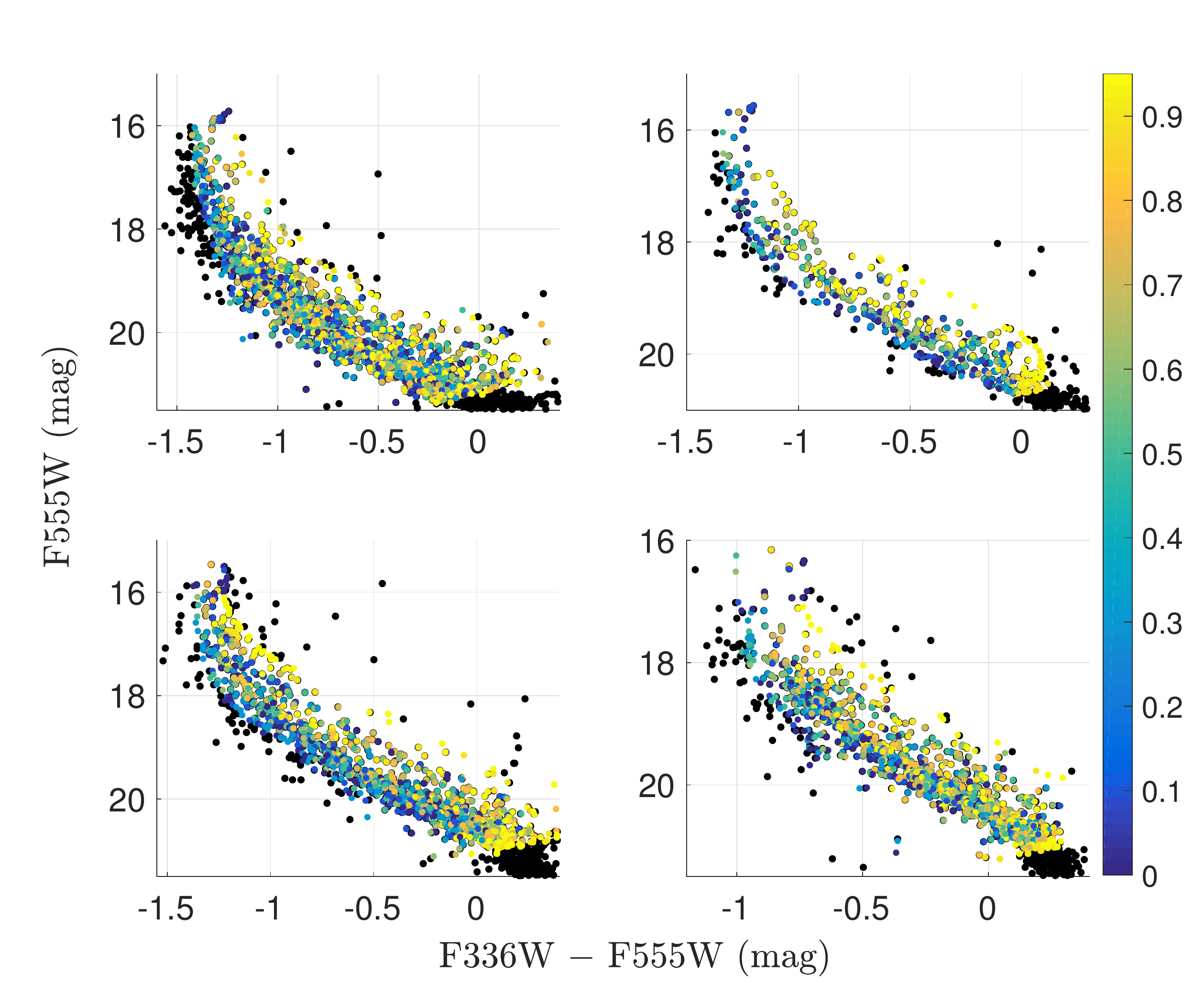}
\caption{Synthetic CMDs for NGC 330, NGC 1805, NGC 1818, and NGC
  2164. The color bar indicates the rotation rates of the
    simulated stars. For each cluster, we adopted 50\% unresolved
  MS--MS binaries. The observed stars are indicated by black dots.}
\label{F4}
\end{figure*}

For each CMD, we generated more than $3\times10^6$ artificial stars. 
Because the Geneva models can only generate rapidly rotating isochrones 
for stellar masses down to 1.7$M_{\odot}$. The numbers of the corresponding 
stars in the CMDs of NGC 330, NGC 1805, NGC 1818, and NGC 2164 
are 1827, 493, 1206, and 1234, respectively. The number of artificial stars is 
thus at least 1600 times larger than the numbers of stars in the 
observations. For each observed star, we selected the 10 nearest 
artificial stars from the synthetic CMD. Finally, we randomly selected 
one of these 10 fake stars as representative of the observed star. Using 
this procedure, if a simple stellar population (SSP) characterized by different 
stellar rotation rates and a realistic fraction of unresolved binaries could
fully cover and match the observed MS region, the synthetic CMD should
be almost identical to the observed CMD. We present our results in
Fig. \ref{F4}.

Fig. \ref{F4} shows that the use of only a coeval stellar population
characterized by different stellar rotation rates cannot fully
reproduce the observed wide MSs. There is an excess of blue MS stars
who cannot be reproduced by our simulations. Our method shows
  that many simulated stars are non-rotating stars (i.e., $\omega$ =
  0.0). This is not surprising, because most fast-rotating stars would
  appear redder than their non-rotating counterparts. Since a fraction
  of the observed MS stars is too blue to be reproduced by our SSP
  models, our method is forced to select numerous non-rotating stars
  to try to cover those blue MS stars in the CMD.
% The fractions of these additional blue MS stars (with magnitudes F555W$\leq19$ mag for NGC 330, NGC 1805, and NGC 1818, and F555W$\leq$20 mag for NGC 2164) are 46.3\%, 45.9\%, 21.9\%, and 32.6\% for NGC 330, NGC 1805, NGC 1818, and NGC 2164, respectively. 
This confirms our speculation based on Fig. \ref{F3}: all sample
clusters exhibit populations of very blue stars that cannot be
explained by a coeval stellar population, not even when adopting a
range in stellar rotation rates. Because the synthetic CMDs have the
same distributions of the photometric uncertainties as the
observations, and the resulting synthetic CMDs are based on large
samples, each containing more than $3\times10^6$ artificial
stars, these additional blue MS stars thus cannot be explained by
large scatter in the measurements. As shown in Fig. \ref{F2}, only few
background stars could contaminate the bright part of the MS; residual
background contamination thus cannot explain these blue MS stars
either. The additional blue stars may suggest the presence of young
stellar populations.

We subsequently explored the age-spread scenario. We used the MIST
models to describe the observations. The MIST models cover a large
grid of single-star stellar evolution models, extending across all
evolutionary phases for different stellar masses and metallicities
\citep{Paxt11a,Paxt13a,Paxt15a,Dott16a,Choi16a}. The MIST 1.0 version
includes a default rotational rate of $V/V_{\rm crit} =0.4
(\omega\sim0.6)$ in the output isochrones. \cite{Huan10a} studied the
rotational velocities of $\sim$530 B-type stars. They found that the
highest probability density occurs around $V/V_{\rm crit}=0.49$, which
is close to the MIST default value. Using the Geneva models, we
confirmed that the color difference between isochrones of $V/V_{\rm
  crit}=0.4$ and 0.49 is small (less than 0.02 mag in $U-V$
color). Given that the broadened sections of our clusters' MSs are
expected to be mainly populated by B-type stars, an average rotational
rate of $V/V_{\rm crit}=0.4 (\omega\sim0.6)$ is thus a good
approximation.

\begin{figure*}
\centering
\includegraphics[width=0.95\textwidth]{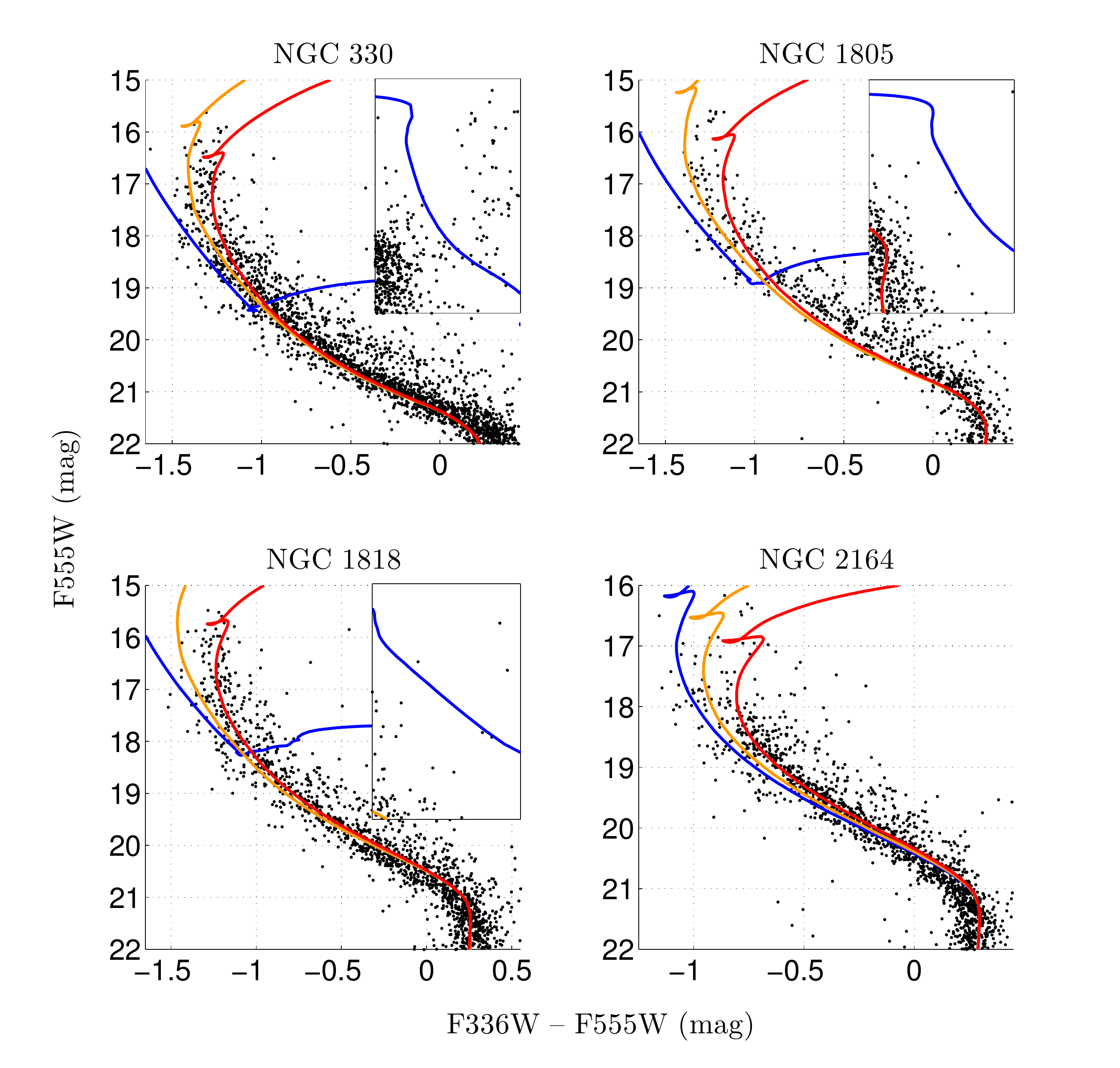}
\caption{CMDs of the clusters NGC 330, NGC 1805, NGC 1818, and NGC
  2164. The blue, orange, and red isochrones represent the loci of
  young, intermediate-age, and old stellar populations. For NGC 330,
  NGC 1805, and NGC 1818, subpanel are included to show the pre-MS
  region down to the bottom of the MS.}
\label{F5}
\end{figure*}

In Fig. \ref{F5} we show the outcome of our fits including age
spreads. For each cluster, we have overplotted a young, a median, and
an old isochrone onto the observations. The best-fitting young and old
isochrones were determined by visual inspection, by comparing their
loci to the blue and red edges of the MSs. The median isochrones were
determined arbitrarily; they will be used to indicate the trends of
the TO regions for different ages. The metallicities, extinction
values, and distance moduli for these differently aged isochrones are
the same: see Table \ref{T3} (rows 5--8).

As shown in Fig. \ref{F5}, as regards our fits to the colors, the
age-spread scenario works better than the stellar rotation
scenario. We found that in order to fit the additional blue MS stars,
the age range of the stellar population needs to extend to almost zero
for NGC 330, NGC 1805, and NGC1818. We simply adopted an age of 1 Myr
to represent the zero-age population in our fits (but see below). For
the relatively old cluster NGC 2164, an isochrone with age down to 50
Myr is required to fit the blue MS boundary. Our fits indicate that
the age spreads in NGC 330, NGC 1805, NGC 1818, and NGC 2164 are 40
Myr, 40 Myr, 35 Myr, and 50 Myr, respectively.

However, the age spread model is not perfect either. The MIST models
suggest that if the age spread were to reflect an extended
star-formation history, a 1 Myr-old stellar population should contain
a fraction of massive O-type and pre-MS stars. However, there is no
obvious evidence of the presence of such objects. In NGC 330, a small
number of stars are detected around the pre-MS locus, but we confirmed
that they are background residuals, based on inspection of their
spatial distribution. The absence of massive O-type and pre-MS stars
indicates that the origin of such young stellar populations must be
reconsidered. We will return to this issue in Section \ref{S4}.

\begin{figure*}
\includegraphics[width=1.0\textwidth]{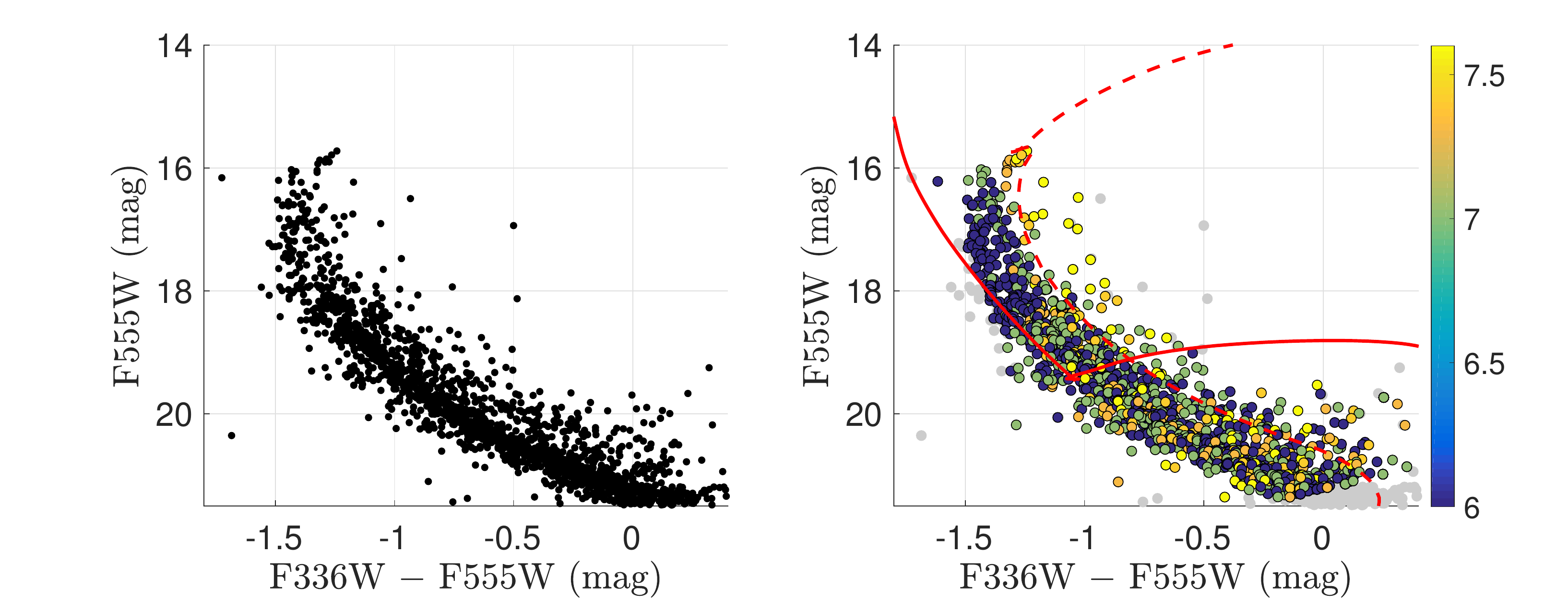}
\caption{(Left) Observed CMD of NGC 330. (Right) Simulated CMD with
  various ages and rotational rates, and observed CMD (grey dots) of
  NGC 330. The color bar represents the ages of the simulated
    stars in the right-hand panel.}
\label{F6}
\end{figure*}

\begin{figure*}
\includegraphics[width=1.0\textwidth]{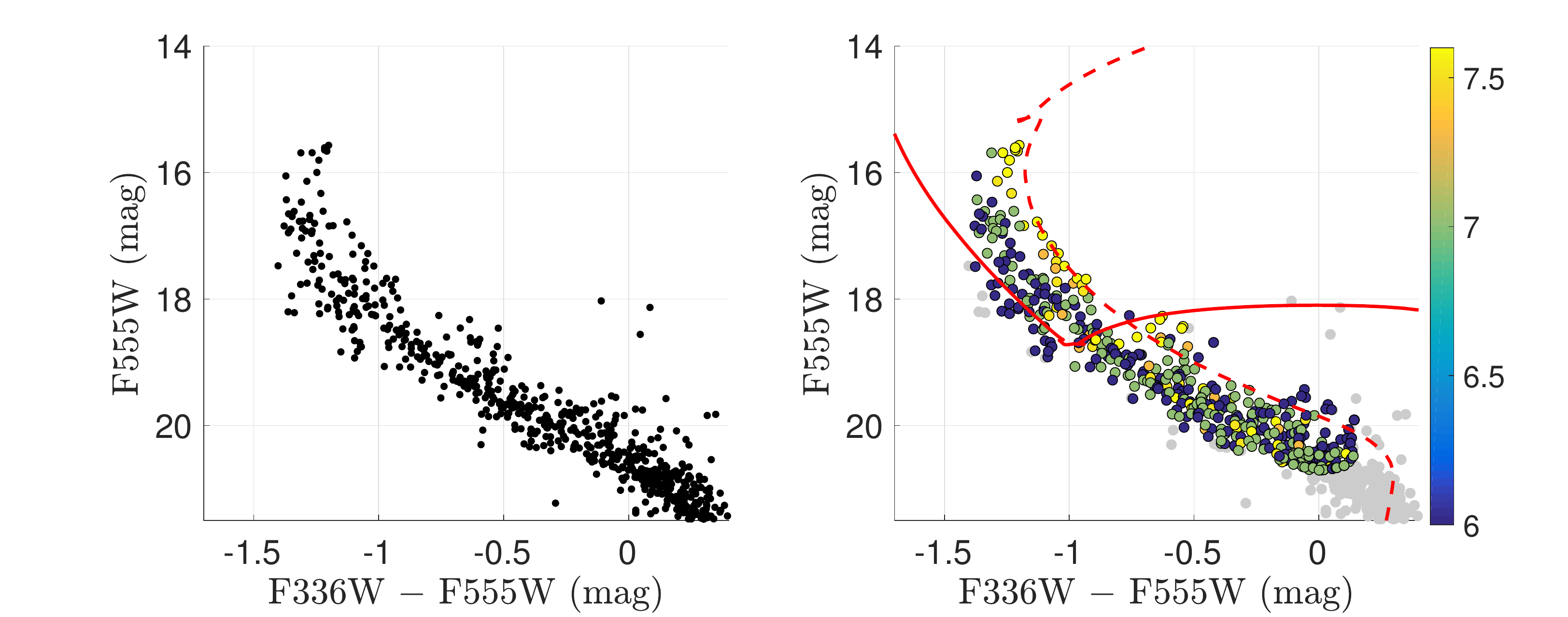}
\caption{As Fig. \ref{F6}, but for NGC 1805.}
\label{F7}
\end{figure*}

\begin{figure*}
\includegraphics[width=1.0\textwidth]{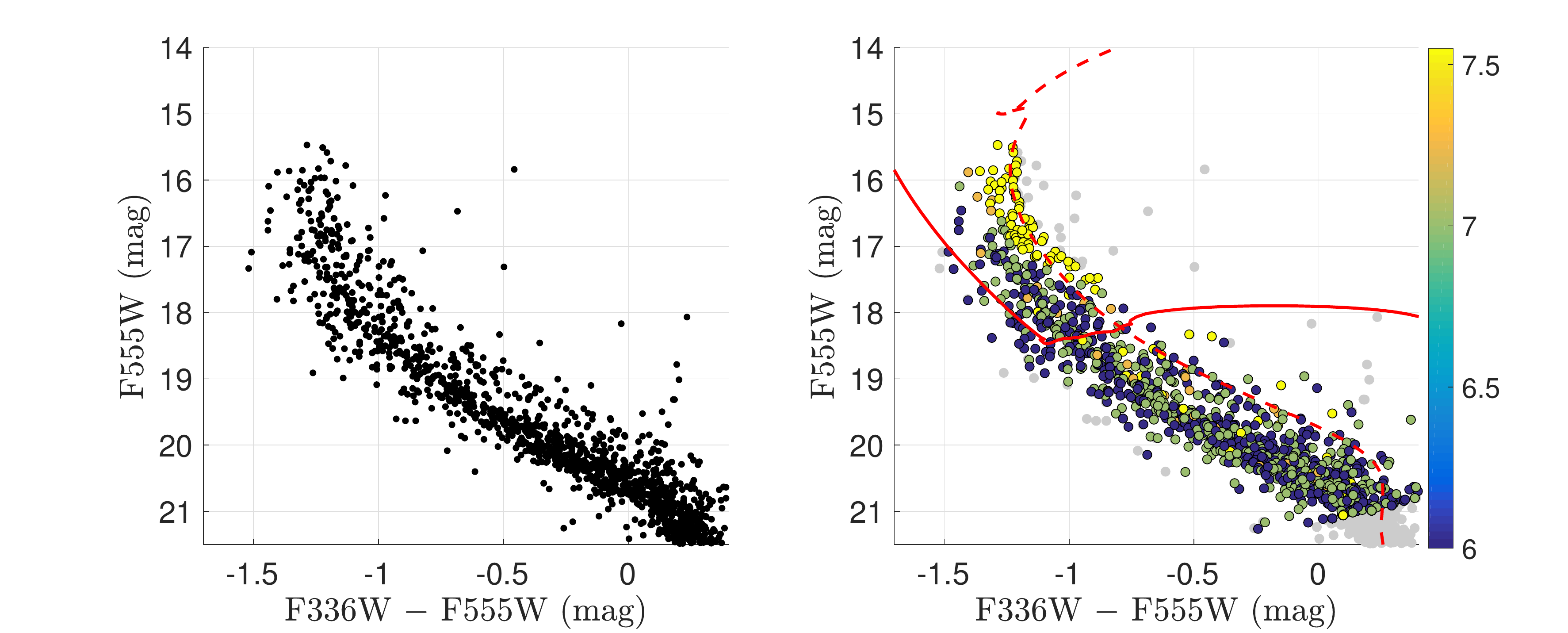}
\caption{As Fig. \ref{F6}, but for NGC 1818.}
\label{F8}
\end{figure*}

\begin{figure*}
\includegraphics[width=1.0\textwidth]{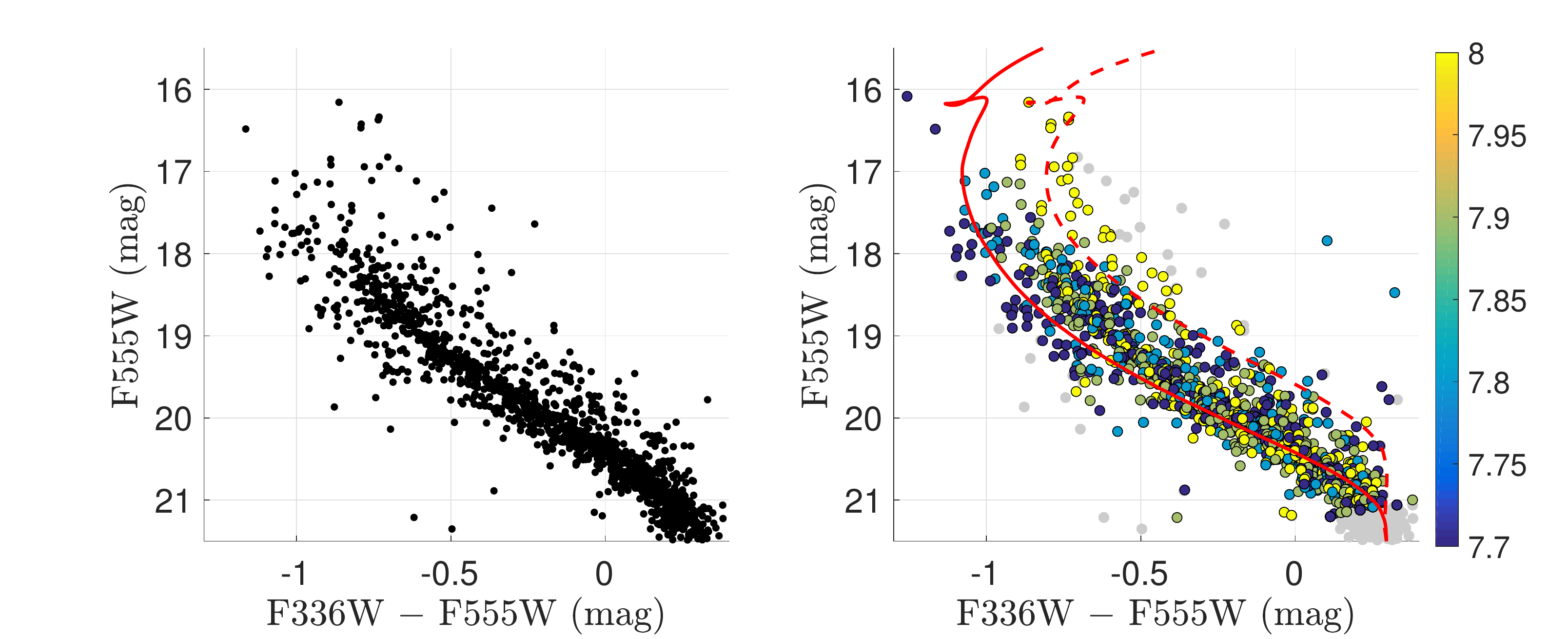}
\caption{As Fig. \ref{F6}, but for NGC 2164.}
\label{F9}
\end{figure*}

Based on their exploration of the TO region in the cluster NGC 1850,
\cite{Corr16a} suggested that the best solution may be a combination
of stellar rotation and an age spread. We similarly explored the
promise of the combination of an age spread and stellar rotation. The
input parameters are presented in Table \ref{T3} (rows 9--12). The
method we used is similar to that used for Fig. \ref{F4}. This time,
we generated artificial stars that are both different in age and
rotation properties. For each observed star, we randomly selected an
artificial star from among the 10 nearest candidates as its best
representation. Our simulation was based on the Geneva database,
because the Geneva models provide isochrones for different stellar
rotation rates. Each time when we generated an SSP for a different
stellar rotation rate, we compared its CMD with the observational
counterpart, similarly to our analysis of Fig. \ref{F4}. Again, we
confirmed that even if we vary the isochronal age, the reproduction is
not satisfactory. Specifically, once we adopted a young isochronal age
for the synthetic CMD, a fraction of red MS stars could not be
reproduced. This is a similar conclusion as that reached by
\citet[][their Fig. 10]{Milo16a}. Synthetic CMDs of stellar
  populations with different ages and rotation rates are presented in
  the right-hand panels of Figs \ref{F6}--\ref{F9}. For a more direct
  comparison, we also included the observed CMDs (gray dots). In the
  synthetic CMDs, colors represent the ages of the simulated stars.

We first examined NGC 330. We found that if we use a combined model
including an age spread and a distribution of stellar rotation rates,
only a small number of blue MS stars would be poorly reproduced.
  We find that for the low-mass end of the main sequence, the
  contributions from young and old populations are not very
  different. However, for the MSTO region, the contributions from the
  young and old populations are indeed very different. Almost all
  young stars reproduce the observed stars located in the blue part of
  the MSTO region, while the old stars mainly contribute to the red
  part of the MSTO. This result further supports the notion that in
  order to reproduce the bluest MS stars, a fraction of young stars is
  required.

The same result for NGC 1805 is shown in Fig. \ref{F7}. The synthetic
CMD of multiply-aged stellar populations is almost identical to
the observations. Only three MS stars around F555W$\sim$18 mag
  are not well reproduced. This is not surprising, because the
synthetic CMD fully covers the region of the observed MS; for each
observed star we should be able to find a corresponding artificial
star with similar color and magnitude. This result stands in
  sharp contrast to the synthetic CMD of a coeval stellar population
  (see Fig. \ref{F4}), where numerous additional blue MS stars in the
  range 16$\leq$F555W$\leq$19 mag are not well reproduced.

This result also holds for NGC 1818, as shown in the right-hand
  panel of Fig. \ref{F8}: almost all blue MS stars that appear in
Fig. \ref{F4} are reproduced. Finally we show the same result for NGC
2164 in Fig. \ref{F9}. Again, the simulated MS is significantly
  broadened compared with that of a single-aged stellar population.
In summary, the cluster CMDs can be reproduced well by a combination
of SSPs that cover an age range of 35--50 Myr and a wide variety of
rotation rates.

At face value, the successful reproduction of the observations through
a combination of an age range and a dispersion in rotation rates seems
to indicate the need for an extended star-formation history for the
clusters. However, the adoption of using a `nearest' star to represent
an observed star is complicated. This treatment is similar to the
application of a mass truncation to the young stellar population,
which thus avoids the appearance of massive O-type stars. This is
indeed a contrived approach.

As we already emphasized, the origin of any ongoing star formation in
these clusters is suspicious. In Figs \ref{F6}--\ref{F9}, we also
included the young isochrones adopted in Fig. \ref{F5}, as well as the
equal-mass binary locus of the corresponding old isochrone. Although
these two lines seem to adequately describe the boundaries of the MSs, the
young isochrone clearly predicts the presence of bright O-type and
pre-MS stars. An absence of O-type stars would indicate an unphysical
mass function, while the lack of pre-MS stars reflects the notion that
star formation has long been terminated.

\section{Discussion}\label{S4}
\subsection{Stellar Rotation?}

We will next discuss the possible physical interpretations of our
observations. As deduced from our analysis of Figs \ref{F3} and
\ref{F4}, the observed wide MSTO regions cannot be fully explained by
a coeval stellar population characterized by different stellar
rotation rates. This is so, because the color separation between the
non-rotating and fast-rotating MS stars is caused by gravity darkening
\citep{von1924a}, which would cause a fast rotator to have a lower
surface temperature than its non-rotating counterpart. Using the
Geneva models, we found that the reduction in surface temperature
caused by a rotation rate of $\omega=0.95$ is roughly constant at
500--1000 K for all MS stars irrespective of their mass. This
temperature difference is sufficient to produce a detectable color
spread for late-B- or F-type stars. However, such a temperature
difference only negligibly affects the colors of more massive, hot
stars. In this paper, most stars located on the bright section of the
MS have typical surface temperatures of $\sim$15,000--20,000 K. The
color difference, $\Delta$(F336W$-$F555W), owing to a reduction of
$\sim$500--1000 K in surface temperature is less than 0.1 mag for
these stars.

The non-rotating and fast-rotating isochrones converge in the MSTO
region, as illustrated by \cite{Gira11a}. This is because stellar
rotation causes expansion of the stellar convective shell, leading to
transportation of shell material to the stellar core. As a result, a
rapidly rotating star will have a longer lifetime during the MS
stage. This process is called rotational mixing. Rotational mixing
renders the MSTO of a fast-rotating population brighter and bluer than
that of a non-rotating population. This effect could therefore mask
the reddening caused by the gravity darkening. 

Because our sample clusters are so young, the observed wide MSs are
mainly composed of early-B-type stars. The surface temperatures of
these massive MS stars is too high for their colors to be
significantly affected by gravity darkening. Rotational mixing causes
populations characterized by different rotation rates to converge into
a narrow sequence at the TO region. All these factors conspire that
stellar rotation cannot fully explain the observed broad MSTOs.

\subsection{Extended Star Formation?}

As we showed in Figs \ref{F6} to \ref{F9}, to reproduce the observed
broad MSTOs, an age spread of 35 Myr (NGC 1818) to 50 Myr (NGC
2164) is required. An age spread of $\sim$10 Myr ($\Delta{\log( t
    \mbox{ yr}^{-1})}\sim$0.1--0.2 for the typical ages of NGC 330,
  NGC 1805, and NGC 1818; roughly equal to the timescale of initial
  gas expulsion) is insufficient. As Figs \ref{F6} to \ref{F9} clearly
  illustrate, only a stellar population with an age as young as 1 Myr
  could reach the positions of the bluest MS stars. Our result is in
agreement with the conclusions of \cite{Milo16b} and \cite{Corr16a}.

Does this imply that all of our sample clusters have extended
star-formation histories? One should be cautious as regards such
speculations. We emphasize once again that the absence of O-type and
pre-MS stars contradicts the ongoing star-formation hypothesis. It
seems that the most massive stars in the blue stellar population are
not more massive than those of the bulk stellar population. As we
showed in Section \ref{S3}, to reproduce the observed MSs, we have to
assume a mass truncation for the young stellar population. This seems
a contrived approach to reproduce the observations, which cannot be
naturally explained by a invoking a scenario of extended star
formation.

A possible explanation is that a few million-year-old stars may still
be embedded in their natal dust cocoons, which would prevent us from
observing them in the UV band. However, this cannot explain the
presence of the very blue MS stars. Why would the very bright O-type
stars be obscured while the B-type stars (and thus those blue MS
stars) are discernible in the UV band? In addition, \cite{Bast14a}
studied the gas and dust contents in NGC 330, NGC 1818, and NGC2164;
no significant gas residuals nor any dust were detected in these
clusters. All these arguments thus challenge the continuous
star-formation hypothesis.

We can estimate the minimum masses for our clusters to retain their
gas using the equation from \cite{Geor09a},
\begin{equation}
M_{\rm cl}\approx100v_{\rm esc}^2r_{\rm h},
\end{equation}
where $M_{\rm cl}$ is the total mass of the cluster expressed in units
of $M_{\odot}$, $v_{\rm esc}$ is the escape velocity of the initial
gas in km s$^{-1}$, and $r_{\rm h}$ is the cluster's half-mass radius
in pc. We assumed that the clusters' half-mass radii are equal to
their half-light radii, $r_{\rm hl}$\footnote{We also adopted an
  average value of 0.1 for the coefficient $f_{\rm c}$ based on table
  2 of \cite{Geor09a}.}. If star formation in these clusters can last
for several tens of millions of years, numerous Type II supernova
explosions should have taken place. The escaping gas can be
accelerated to several hundred km s$^{-1}$. The corresponding minimum
masses required to retain the initial gas in our sample clusters are
log$(M_{\rm cl}/M_{\odot})$ = 6.84, 6.46, 6.70, and 6.64 (assuming a
minimum escape velocity of 100 km s$^{-1}$). However, the current
masses of NGC 330, NGC 1805, NGC 1818, and NGC2164 are only
log$(M_{\rm cl}/M_{\odot})$ = 4.61, 3.70, 4.41, and 4.18, respectively
\citep{Mcla05a}. Clearly, it is difficult for these clusters to
sustain long star-formation episodes by accreting their initial gas.

Another explanation to explain the occurrence of an extended
star-formation episode that seems viable is that after the initial gas
expulsion phase ($\sim$10 Myr), these clusters may still have been
sufficiently massive to accrete the subsequent stellar ejecta of the
asymptotic giant-branch (AGB) stars. The minimum velocity for the AGB
stellar ejecta is about $v_{\rm esc}$ = 10 km s$^{-1}$
\citep{Renz08a,Li16c}. The minimum masses to retain the gas of the AGB
ejecta for these clusters are log$(M_{\rm /}{M}_{\odot})$ = 4.84,
4.46, 4.70, and 4.64, which is still 1.7 (NGC 330) to 5.8 (NGC 1805)
times the current cluster masses. Although the clusters would have
lost their stellar mass through dynamical evaporation, this would not
dramatically affect our clusters, because the typical timescale for
such mass loss is expressed in units of billions of years
\citep{Mcla08a,Li16c}.

In summary, although the observed broad MSs in our clusters indicate
the possible presence of an age spread, it seems that such age spreads
unlikely originate from extended star-formation histories.

\subsection{Blue Straggler Stars?}

An alternative explanation is that the population of puzzling blue MS
stars are blue straggler stars (BSSs). Because BSSs are produced
through binary mass transfer, mergers, or stellar collisions, their
maximum mass does not exceed twice the mass of TO stars. This may
explain the absence of O-type and pre-MS stars.

\cite{Dant15a} studied the split MS in the cluster NGC 1856. They
speculated that all observed blue MS-component stars may hide a binary
component. They suggested that the periods pertaining to these binary
systems range from 4 to 500 days. Based on Kepler's Third Law, the
relationship between the binary period, $P$, and its semi-major axis,
$a$, is
\begin{equation}\label{eq3}
P=2\pi\sqrt{\frac{a^3}{G(M+m)}}=2\pi\sqrt{\frac{a^3}{GM(1+q)}},
\end{equation}
where $G$ is the usual gravitational constant, $M$ and $m$ are the
masses of the primary and secondary stars, and $q=m/M$ is their mass
ratio. Based on equation \ref{eq3}, we can calculate $a$ in terms of
$P$, $M$, and $q$:
\begin{equation}\label{eq3}
a=\sqrt[3]{\frac{P^2GM(1+q)}{4\pi^2}}.
\end{equation}
Assuming that $q$ ranges from 0 to 1, $P$ from 4 to 500 days, and that
the typical mass of the young-population stars in our clusters is
2$M_{\odot}\leq{M}\leq9M_{\odot}$, the resulting distribution of the
semi-major axes ranges from from $a = 0.06$ to 3.23 au. Using a Monte
Carlo approach, we derived that the average length of the semi-major
axisof the binary population is $a = 1.63$ au, with 23\% of the
binaries having semi-major axes $a\leq$1 au (we adopted flat
distributions for $q$, $P$, and $M$). This indicates the presence of a
significant fraction of compact binaries, which may form a potential
population for mass-transfer BSS candidates. This was also suggested
by \cite{Yang11a}. Specifically, the latter authors suggested that the
eMSTO regions in intermediate-age star clusters include significant
contributions from binary interactions and mergers.

However, the fraction of BSSs with respect to the bulk stellar
population is usually very small \citep[e.g., for the old LMC GCs,
  see][]{Li13b,Mack06a}. It is unclear how a cluster environment could
produce such a large number of BSSs. To our knowledge, there are no
direct observations of BSSs in YMCs. \cite{Xin07a} studied a large
sample of ($> 1$ Gyr-old) Galactic open clusters. They found that the
specific frequency of BSSs (i.e., the ratio of the number of BSSs to
that of stars spanning a 2 mag range below the MSTO) ranges from 1\%
to 20\%. Numerical simulations also show that the specific frequency
of BSSs should be small during the early stages of a cluster's
evolution \citep[e.g., at an age of 50 Myr, the number of BSSs is
  fewer than 50 for a cluster that initially contains 100,000
  stars;][]{Lu11a,Hypk13a}.

Despite this challenge, if the observed young stars are indeed BSSs,
most should have been formed through binary mass transfer. Future
studies should focus on their carbon and oxygen abundances to test
this suggestion \citep{Ferr06a}.

\section{Conclusions}\label{S5}

In this paper, we have studied the CMDs of the clusters NGC 330, NGC
1805, NGC 1818, and NGC 2164. They all exhibit broad MSs, which cannot
be explained by a SSP with unresolved binaries and photometric
uncertainties. We suggest that it is likely that most YMCs may exhibit
wide MSs in their ultraviolet--visual CMDs.

We found that the gravity darkening caused by stellar rotation plays a
very limited role in hot, massive, early-type MS stars. In the
meantime, rotational mixing would cause an isochrone to have a TO
position that is almost indistinguishable from that of a non-rotating
isochrone. Therefore, a dispersion of stellar rotation velocities in
coeval ensembles of stars cannot reproduce the observed wide MSs.

The failure of the stellar rotation scenario to fully reproduce the
observations indicates that the age-spread scenario may still be
viable. An age spread of 35--50 Myr is required to explain the
observations. We confirm that a combination of an age spread and
stellar rotation can adequately reproduce the observations for all
sample clusters. Similar conclusion were derived previously for the
$\sim$100--200-Myr old clusters NGC 1850 and NGC 1866
\citep{Milo16b,Corr16a}. However, we argue that the apparent age
spread is unlikely owing to continuous star formation. Indeed, the
clusters' masses seem too small to substain extended star-formation
episodes. Moreover, the absence of O-type and pre-MS stars also
contradicts the extended star-formation hypothesis.

We suggest that the young stars may be BSSs, which reduces the need
for very massive O-type and pre-MS stars. However, if this were
correct, it is not clear why the number of BSSs in the clusters should
be comparable to the number of the bulk stellar population. To
understand the origin of these young stars, details about their
chemical composition and rotation rates are required.

\begin{acknowledgments} 
We thank Choi Jieun (Harvard--Smithsonian Center for Astrophysics) for
providing us with the relevant model database. C. L. is supported by
the Macquarie Research Fellowship Scheme. R. d. G. and
L. D. acknowledge research support from the National Natural Science
Foundation of China through grants U1631102, 11373010, and
11633005. A. P. M. acknowledges support from the Australian Research
Council through a Discovery Early Career Researcher Award, number
DE150101816.
\end{acknowledgments}

\bibliographystyle{apj}
%\bibliography{reference}

\begin{thebibliography}{33}
\expandafter\ifx\csname
natexlab\endcsname\relax\def\natexlab#1{#1}\fi
\expandafter\ifx\csname natexlab\endcsname\relax\def\natexlab#1{#1}\fi

\bibitem[Bastian \& de Mink(2009)]{Bast09a} Bastian, N., \& de Mink,
  S.~E.\ 2009, \mnras, 398, L11

\bibitem[Bastian \& Strader(2014)]{Bast14a} Bastian, N., \& Strader,
  J.\ 2014, \mnras, 443, 3594

\bibitem[Bastian \& Niederhofer(2015)]{Bast15a} Bastian, N., \&
  Niederhofer, F.\ 2015, \mnras, 448, 1863

\bibitem[Bastian et al.(2016)]{Bast16a} Bastian, N., Niederhofer, F.,
  Kozhurina-Platais, V., et al.\ 2016, \mnras, 460, L20

\bibitem[Bertelli et al.(2003)]{Bert03a} Bertelli, G., Nasi, E.,
  Girardi, L., et al.\ 2003, \aj, 125, 770

\bibitem[Bessell(1991)]{Bess91a} Bessell, M.~S.\ 1991, in: The
  Magellanic Clouds, Proc. IAU Symp. 148, Haynes, R., \& Milne, D.,
  eds, Dordrecht: Kluwer, p. 273

\bibitem[Brandt \& Huang(2015)]{Bran15a} Brandt, T.~D., \& Huang,
  C.~X.\ 2015, \apj, 807, 24

\bibitem[Castelli \& Kurucz(2004)]{Cast04a} Castelli, F., \& Kurucz,
  R.~L.\ 2004, in: Modelling of Stellar Atmospheres, Proc. IAU
  Symp. 210, Poster A20, Piskunov, N., Weiss, W. W., \& Gray, D. F.,
  eds, San Francisco: ASP; arXiv:astro-ph/0405087

\bibitem[Choi et al.(2016)]{Choi16a} Choi, J., Dotter, A., Conroy, C.,
  et al.\ 2016, \apj, 823, 102

\bibitem[Correnti et al.(2017)]{Corr16a} Correnti, M., Goudfrooij, P., Bellini, A., Kalirai, J.~S., \& Puzia, T.~H.\ 2017, \mnras, 467, 3628 

\bibitem[D'Antona et al.(2015)]{Dant15a} D'Antona, F., Di Criscienzo,
  M., Decressin, T., et al.\ 2015, \mnras, 453, 2637

\bibitem[de Grijs et al.(2014)]{grijs14a} de Grijs, R., Wicker, J.~E.,
  \& Bono, G.\ 2014, \aj, 147, 122

\bibitem[de Grijs \& Bono(2015)]{grijs15a} de Grijs, R., \& Bono,
  G.\ 2015, \aj, 149, 179

\bibitem[de Grijs(2017)]{grijs17a} de Grijs, R.\ 2017, Nat. Astron.,
  1, 0011

\bibitem[de Mink et al.(2013)]{Demi13a} de Mink, S.~E., Langer, N.,
  Izzard, R.~G., Sana, H., \& de Koter, A.\ 2013, \apj, 764, 166

\bibitem[Dolphin(2011a)]{Dolp11a} Dolphin A., DOLPHOT/WFC3 user's
  guide, version 2.0. Available from {\url
    {http://americano.dolphinsim.com/dolphin/dolphotWFC3.pdf}}

\bibitem[Dolphin(2011b)]{Dolp11b} Dolphin A., DOLPHOT/WFPC2 user's
  guide, version 2.0. Available from {\url
    {http://americano.dolphinsim.com/dolphot/dolphotWFPC2.pdf}}

\bibitem[Dolphin(2013)]{Dolp13a} Dolphin A., DOLPHOT user's guide,
  version 2.0. Available from {\url
    {http://americano.dolphinsim.com/dolphot/dolphot.pdf}}

\bibitem[Dotter(2016)]{Dott16a} Dotter, A.\ 2016, \apjs, 222, 8

\bibitem[Ekstr{\"o}m et al.(2012)]{Ekst12a} Ekstr{\"o}m, S., Georgy,
  C., Eggenberger, P., et al.\ 2012, \aap, 537, A146

\bibitem[Ferraro et al.(2006)]{Ferr06a} Ferraro, F.~R., Sabbi, E.,
  Gratton, R., et al.\ 2006, \apjl, 647, L53

\bibitem[Georgiev et al.(2009)]{Geor09a} Georgiev, I.~Y., Hilker, M.,
  Puzia, T.~H., Goudfrooij, P., \& Baumgardt, H.\ 2009, \mnras, 396,
  1075

\bibitem[Georgy et al.(2013)]{Geor13a} Georgy, C., Ekstr{\"o}m, S.,
  Granada, A., et al.\ 2013, \aap, 553, A24

\bibitem[Georgy et al.(2014)]{Geor14a} Georgy, C., Granada, A.,
  Ekstr{\"o}m, S., et al.\ 2014, \aap, 566, A21

\bibitem[Girardi et al.(2009)]{Gira09a} Girardi, L., Rubele, S., \&
  Kerber, L.\ 2009, \mnras, 394, L74

\bibitem[Girardi et al.(2011)]{Gira11a} Girardi, L., Eggenberger, P.,
  \& Miglio, A.\ 2011, \mnras, 412, L103

\bibitem[Girardi et al.(2013)]{Gira13a} Girardi, L., Goudfrooij, P.,
  Kalirai, J.~S., et al.\ 2013, \mnras, 431, 3501

\bibitem[Girardi(2016)]{Gira16a} Girardi, L.\ 2016, \araa, 54, 95

\bibitem[Goudfrooij et al.(2011)]{Goud11a} Goudfrooij, P., Puzia,
  T.~H., Kozhurina-Platais, V., \& Chandar, R.\ 2011, \apj, 737, 3

\bibitem[Goudfrooij et al.(2014)]{Goud14a} Goudfrooij, P., Girardi,
  L., Kozhurina-Platais, V., et al.\ 2014, \apj, 797, 35

\bibitem[Gratton et al.(2004)]{Grat04a} Gratton, R., Sneden, C., \&
  Carretta, E.\ 2004, \araa, 42, 385

\bibitem[Huang et al.(2010)]{Huan10a} Huang, W., Gies, D.~R., \&
  McSwain, M.~V.\ 2010, \apj, 722, 605

\bibitem[Hypki \& Giersz(2013)]{Hypk13a} Hypki, A., \& Giersz,
  M.\ 2013, \mnras, 429, 1221

\bibitem[Johnson et al.(2001)]{John01a} Johnson, R.~A., Beaulieu,
  S.~F., Gilmore, G.~F., et al.\ 2001, \mnras, 324, 367

\bibitem[King(1966)]{King66a} King, I.~R.\ 1966, \aj, 71, 64

\bibitem[Krause et al.(2016)]{Krau16a} Krause, M.~G.~H., Charbonnel,
  C., Bastian, N., \& Diehl, R.\ 2016, \aap, 587, A53

\bibitem[Li et al.(2013a)]{Li13a} Li, C., de Grijs, R., \& Deng,
  L.\ 2013a, \mnras, 436, 1497

\bibitem[Li et al.(2013b)]{Li13b} Li, C., de Grijs, R., Deng, L., \&
  Liu, X.\ 2013b, \apjl, 770, L7

\bibitem[Li et al.(2014a)]{Li14a} Li, C., de Grijs, R., \& Deng,
  L.\ 2014a, \apj, 784, 157

\bibitem[Li et al.(2014b)]{Li14b} Li, C., de Grijs, R., \& Deng,
  L.\ 2014b, \nat, 516, 367

\bibitem[Li et al.(2016a)]{Li16a} Li, C., de Grijs, R., Bastian, N.,
  et al.\ 2016a, \mnras, 461, 3212

\bibitem[Li et al.(2016b)]{Li16c} Li, C., de Grijs, R., \& Deng,
  L.\ 2016b, RAA, 16, 179

\bibitem[Li et al.(2017)]{Li16b} Li, C., de Grijs, R., Deng, L., \&
  Milone, A.~P.\ 2017, \apj, 834, 156

\bibitem[Lu et al.(2011)]{Lu11a} Lu, P., Deng, L.-C., \& Zhang,
  X.-B.\ 2011, RAA, 11, 1336

\bibitem[Lupton et al.(1989)]{Lupt89a} Lupton, R.~H., Fall, S.~M.,
  Freeman, K.~C., \& Elson, R.~A.~W.\ 1989, \apj, 347, 201

\bibitem[Mackey et al.(2006)]{Mack06a} Mackey, A.~D., Payne, M.~J., \&
  Gilmore, G.~F.\ 2006, \mnras, 369, 921

\bibitem[Mackey \& Broby Nielsen(2007)]{Mack07a} Mackey, A.~D., \&
  Broby Nielsen, P.\ 2007, \mnras, 379, 151

\bibitem[McLaughlin \& van der Marel(2005)]{Mcla05a} McLaughlin,
  D.~E., \& van der Marel, R.~P.\ 2005, \apjs, 161, 304

\bibitem[McLaughlin \& Fall(2008)]{Mcla08a} McLaughlin, D.~E., \&
  Fall, S.~M.\ 2008, \apj, 679, 1272-1287

\bibitem[Milone et al.(2009)]{Milo09a} Milone, A.~P., Bedin, L.~R.,
  Piotto, G., \& Anderson, J.\ 2009, \aap, 497, 755

\bibitem[Milone et al.(2012)]{Milo12a} Milone, A.~P., Piotto, G.,
  Bedin, L.~R., et al.\ 2012, \aap, 540, A16

\bibitem[Milone et al.(2015)]{Milo15a} Milone, A.~P., Bedin, L.~R.,
  Piotto, G., et al.\ 2015, \mnras, 450, 3750

\bibitem[Milone et al.(2016)]{Milo16a} Milone, A.~P., Marino, A.~F.,
  D'Antona, F., et al.\ 2016, \mnras, 458, 4368

\bibitem[Milone et al.(2017)]{Milo16b} Milone, A.~P., Marino, A.~F.,
  D'Antona, F., et al.\ 2017, \mnras, 465, 4363

\bibitem[Mucciarelli et al.(2006)]{Mucc06a} Mucciarelli, A., Origlia,
  L., Ferraro, F.~R., Maraston, C., \& Testa, V.\ 2006, \apj, 646, 939

\bibitem[Niederhofer et al.(2017)]{Nied16a} Niederhofer, F., Bastian,
  N., Kozhurina-Platais, V., et al.\ 2017, \mnras, 465, 4159

\bibitem[Piotto et al.(2015)]{Piot15a} Piotto, G., Milone, A.~P.,
  Bedin, L.~R., et al.\ 2015, \aj, 149, 91

\bibitem[Paxton et al.(2011)]{Paxt11a} Paxton, B., Bildsten, L.,
  Dotter, A., et al.\ 2011, \apjs, 192, 3

\bibitem[Paxton et al.(2013)]{Paxt13a} Paxton, B., Cantiello, M.,
  Arras, P., et al.\ 2013, \apjs, 208, 4

\bibitem[Paxton et al.(2015)]{Paxt15a} Paxton, B., Marchant, P.,
  Schwab, J., et al.\ 2015, \apjs, 220, 15

\bibitem[Renzini(2008)]{Renz08a} Renzini, A.\ 2008, \mnras, 391, 354

\bibitem[Sagar \& Richtler(1991)]{Saga91a} Sagar, R., \& Richtler,
  T.\ 1991, \aap, 250, 324

\bibitem[Salinas et al.(2016)]{Sali16a} Salinas, R., Pajkos, M.~A.,
  Strader, J., Vivas, A.~K., \& Contreras Ramos, R.\ 2016, \apjl, 832,
  L14

\bibitem[Sarna \& De Greve(1996)]{Sarn96a} Sarna, M.~J., \& De Greve,
  J.-P.\ 1996, \qjras, 37, 11

\bibitem[Scowcroft et al.(2016)]{Scow16a} Scowcroft, V., Freedman,
  W.~L., Madore, B.~F., et al.\ 2016, \apj, 816, 49

\bibitem[Vallenari et al.(1991)]{Vall91a} Vallenari, A., Chiosi, C.,
  Bertelli, G., Meylan, G., \& Ortolani, S.\ 1991, \aaps, 87, 517

\bibitem[von Zeipel(1924)]{von1924a} von Zeipel, H.\ 1924, \mnras, 84,
  665

\bibitem[Wilson(1975)]{Wils75a} Wilson, C.~P.\ 1975, \aj, 80, 175

\bibitem[Wu et al.(2016)]{Wu16a} Wu, X., Li, C., de Grijs, R., \&
  Deng, L.\ 2016, \apjl, 826, L14

\bibitem[Xin et al.(2007)]{Xin07a} Xin, Y., Deng, L., \& Han,
  Z.~W.\ 2007, \apj, 660, 319

\bibitem[Yang et al.(2011)]{Yang11a} Yang, W., Meng, X., Bi, S., et
  al.\ 2011, \apjl, 731, L37

\bibitem[Yang et al.(2013)]{Yang13a} Yang, W., Bi, S., Meng, X., \&
  Liu, Z.\ 2013, \apj, 776, 112

\end{thebibliography}

%\begin{thebibliography}{}
%\end{thebibliography}

\end{document}